\documentclass[aps,twocolumn,showpacs,showkeys,superscriptaddress,amsmath,amssymb,nofootinbib,floatfix,prc]{revtex4-1}

\usepackage{graphicx}
\usepackage{subfigure}
\usepackage{bm}

\makeatletter

\begin{document}

\title{Torqued fireballs in relativistic heavy-ion collisions}

\author{Piotr Bo\.zek}
\email{Piotr.Bozek@ifj.edu.pl}
\affiliation{The H. Niewodnicza\'nski Institute of Nuclear Physics, Polish Academy of Sciences, PL-31342 Krak\'ow, Poland}
\affiliation{Institute of Physics, Rzesz\'ow University, PL-35959 Rzesz\,ow, Poland}

\author{Wojciech Broniowski}
\email{Wojciech.Broniowski@ifj.edu.pl} 
\affiliation{The H. Niewodnicza\'nski Institute of Nuclear Physics, Polish Academy of Sciences, PL-31342 Krak\'ow, Poland} 
\affiliation{Institute of Physics, Jan Kochanowski University, PL-25406~Kielce, Poland}

\author{Jo\~{a}o Moreira}
\email{jmoreira@lca.uc.pt} 
\affiliation{Centro de F\'{i}sica Computacional, Department of Physics, University of Coimbra, 3004-516 Coimbra, Portugal}

\date{15 November 2010}

\begin{abstract}
We show that the fluctuations in the wounded-nucleon model of the initial stage of relativistic heavy-ion
collisions, together with the natural assumption that the forward (backward) moving wounded nucleons 
emit particles preferably in the forward (backward) direction, lead to an event-by-event 
torqued fireball. The principal axes associated with the transverse shape are rotated in the forward 
region in the opposite direction than in the backward region. On the average, the 
standard deviation of the relative torque angle between the forward 
and backward rapidity regions is about $20^\circ$ for the central and $10^\circ$ for the mid-peripheral collisions.
The hydrodynamic expansion of a torqued fireball leads to a torqued 
collective flow, yielding, in turn,
 torqued principal axes of the transverse-momentum
distributions at different rapidities.
We propose experimental measures,
 based on cumulants involving particles in different rapidity regions, 
which should allow for a quantitative determination of the effect from the data. 
To estimate the non-flow contributions from resonance decays we run Monte Carlo simulations with {\tt THERMINATOR}.
If the event-by-event torque effect is found in the data, it will support the assumptions concerning the 
fluctuations in the early stage of the fireball 
formation, as well as the hypothesis of the asymmetric rapidity shape of the emission functions 
of the moving sources in the nucleus-nucleus collisions.   
\end{abstract}



\maketitle

\section{Introduction \label{sec:intro}}

It is believed that the forward-backward (FB) rapidity correlations 
may reveal important information on the mechanism of particle 
production in high-energy hadronic and nuclear collisions. 
Long-range rapidity correlations uncover properties
of the dynamical system  at a very early stage.
For that reason the FB multiplicity  
correlations have been studied experimentally 
\cite{Uhlig:1977dc,*Alpgard:1983xp,*Aivazian:1988ui,*Steinberg:2005ec,*Tarnowsky:2006nh,*Srivastava:2007ei,*:2009qa,*:2009dqa,Back:2006id} and theoretically 
\cite{Benecke:1976fd,*Capella:1982ru,*Fialkowski:1982kt,*DiasdeDeus:1987bn,*Shuryak:2000pd,*Brogueira:2006yk,*Brogueira:2007ub,*Haussler:2006rg,*Cunqueiro:2006xe,*Armesto:2006bv,*Koch:2008ia,*Konchakovski:2008cf,*Gelis:2009tg,*Dusling:2009ni,*Brogueira:2009nj,Bialas:2004su,Bialas:2004kt,*Fialkowski:2004wh,*Brodsky:1977de,*Adil:2005qn,*Adil:2005bb,Gazdzicki:2005rr,*Bzdak:2009xq,*Bialas:2010zb,Bzdak:2009dr,Bozek:2010bi}. 
In this paper we show that the wounded-nucleon approach \cite{Bialas:1976ed,Bialas:2008zza} to the initial stage of 
the collisions leads to a new manifestation of the FB fluctuations: the {\em torqued fireball}. 
Specifically,  parts of the fireball are rotated in the transverse 
plane in one direction in the forward rapidity region, and in the opposite direction in the backward rapidity region.
The  magnitude and sign of the torque angle  fluctuate from event to event. 

The following ingredients are responsible for the appearance of the torque effect: 
\begin{enumerate}
\item statistical fluctuations of the transverse density of the sources
(wounded nucleons) 
\cite{Aguiar:2001ac,*Manly:2005zy,*Andrade:2006yh,*Voloshin:2006gz,*Alver:2006wh,*Drescher:2006ca,*Broniowski:2007ft,*Voloshin:2007pc,*Hama:2009pk,*Andrade:2008fa,*Broniowski:2009fm}, and 
\item the asymmetric shape \cite{Bialas:2004su,Bialas:2004kt,Gazdzicki:2005rr,Bzdak:2009dr,Bozek:2010bi}
of the particle emission function, peaked in the forward (backward) rapidity for the 
forward (backward) moving wounded nucleons. 
\end{enumerate}

On the average, for the studied Au+Au collisions at 
the highest energies available at the 
BNL Relativistic Heavy Ion Collider 
(RHIC), 
the relative torque angle between the principal axes in the forward 
and backward rapidity regions is about $20^\circ$ for the central, and $10^\circ$ for the mid-peripheral collisions.
On the other hand, an experimental observation of nonzero torque angles
of the expanding fireball could shed light on the mechanism of the formation
of dense matter. A finite torque between reaction planes at different rapidities could influence  elliptic and directed flow studies using reaction planes determined in different pseudorapidity intervals \cite{Adams:2004bi,*Afanasiev:2009wq}.  

The paper has three parts, referring subsequently to the early wounded-nucleon phase of the collision (Sect.~\ref{sec:torqued}), 
the intermediate hydrodynamic stage (Sect.~\ref{sec:hydro}), and, finally, the 
statistical hadronization phase (Sect.~\ref{sec:exp}), 
where hadrons are produced. 
In Sect.~\ref{sec:grad} we introduce the necessary elements 
of the wounded-nucleon approach, in particular, the asymmetric rapidity-dependent emission functions \cite{Bialas:2004su,Bialas:2004kt,Gazdzicki:2005rr,Bzdak:2009dr,Bozek:2010bi} of the forward 
and backward moving wounded nucleons. In Sect.~\ref{sec:fluct} we explain how statistical fluctuations of the 
density of the forward and backward moving wounded nucleons in the transverse plane lead to the event-by-event torque effect. 
Simulations are carried out with {\tt GLISSANDO} \cite{Broniowski:2007nz} in the so-called mixed model 
\cite{Kharzeev:2000ph,Back:2001xy,*Back:2004dy}, 
incorporating an admixture of binary collisions into the wounded-nucleon model.
We introduce quantitative measures of the torque in Sec.~\ref{sec:mtf}. Specifically, 
we use the difference of the forward and backward 
angles of the principal axes, as well as three angles, forward, backward, and central, to construct
useful statistical quantities. 
Next, in Sect.~\ref{sec:hydro} we present the results of running the 3+1 dimensional (perfect fluid)
hydrodynamics, as described in~\cite{Bozek:2009ty}. The calculations show that the torque survives the hydrodynamic stage of the evolution.
Then, in Sect.~\ref{sec:stat} we pass to discussing the final stage, namely,
 the statistical hadronization
(for a review, see, e.g. \cite{Florkowski:2010zz}). 
This stage would wash out the torque effect, 
unless careful experimental measures are used. This is because even at a fixed spatial geometry of the fireball the finiteness of the 
number of produced particles causes large fluctuations of the event-plane angle, covering up the torque angle. 
We thus propose to investigate measures based on cumulants \cite{Ollitrault:1997vz}, 
introduced in Sect.~\ref{sec:cumul}. They are constructed in such a way that the relative FB torque angle 
can be extracted. We examine the non-flow contributions to these measures via
simulations in {\tt THERMINATOR} \cite{Kisiel:2005hn}, proving that 
it is possible to find 
the torque effect  with the proposed methods 
in the large-statistics RHIC data. 
In Sect.~\ref{sec:fbc} we repeat this analysis for the particles taken from three rapidity bins, forward, backward, and central, 
with similar conclusions. 

\section{Torqued fireball \label{sec:torqued}}

In this Section we describe the earliest stage of the relativistic heavy-ion collision in terms of the wounded-nucleon 
model \cite{Bialas:1976ed,Bialas:2008zza}. All simulations are carried out for the Au+Au collisions at the highest RHIC energy of 
$\sqrt{s_{NN}}=200$~GeV.

\subsection{Wounded nucleons with rapidity profiles \label{sec:grad}}

The wounded-nucleon \cite{Bialas:1976ed,*Bialas:2008zza} approach is commonly used to describe the early stage of the heavy-ion collisions. 
In the collision process $N_w$ nucleons get wounded, i.e. interact inelastically at least once, 
as well as $N_{\rm bin}$ binary collisions occur. 
Roughly speaking, wounded nucleon (participants) are responsible for the soft emission, while 
binary collisions describe hard processes. Both serve 
as {\em sources} for the formation of the density of energy or entropy in the initial fireball. 
At RHIC, the mixed model \cite{Kharzeev:2000ph}, where the total number of the produced particles is given by the combination 
\begin{eqnarray}
N_{\rm prod} = A \left ( \frac{1-\alpha}{2} N_w+\alpha N_{\rm bin} \right ),
\end{eqnarray}
describes quite successfully the multiplicity data \cite{Back:2001xy,*Back:2004dy}. The proportionality constant $A$ depends on the energy 
of the collision but not on its centrality. Throughout this work we use $\alpha=0.145$, corresponding 
to collisions at $\sqrt{s_{NN}}=200$~GeV \cite{Back:2001xy,*Back:2004dy}.

The concept of sources can be extended to account for the rapidity dependence of the produced 
particles. The pseudorapidity distribution is given as a sum
of contributions from the emission of forward and backward going wounded
nucleons. Within such a  scheme
Bia\l{}as and Czy\.z successfully described \cite{Bialas:2004su} the distribution of charged particles in pseudorapidity in 
the deuteron-gold collisions
 \cite{Nouicer:2004ke}. The independent emission from the forward and backward going nucleons determines specific FB multiplicity correlations 
 \cite{Gazdzicki:2005rr,Bzdak:2009dr}. In particular, this
 hypothesis  has been tested 
successfully in Ref.~\cite{Bzdak:2009dr} with the FB multiplicity
 correlation data from the PHOBOS collaboration \cite{Back:2006id}. 
Based on this idea, 
Ref. \cite{Bozek:2010bi} postulated that  in nucleus-nucleus 
collisions the emission profile defining the initial density 
(in the space-time rapidity $\eta_\parallel$ and the transverse plane coordinates $x$, $y$) 
has the form 
\begin{eqnarray}
F(\eta_\parallel,x,y)&=&(1-\alpha)[\rho_+(x,y) f_+(\eta_\parallel)
+ \rho_-(x,y) f_-(\eta_\parallel)] \nonumber \\
&+&\alpha 
\rho_{\rm bin}(x,y) f(\eta_\parallel), 
\label{eq:em}
\end{eqnarray}
where $\rho_\pm(x,y)$ is the density of the forward and backward going 
wounded nucleons at a given point in the 
transverse plane, $\rho_{bin}(x,y)$ is the binary collisions density, 
while $f_+(\eta_\parallel)$  and $f_-(\eta_\parallel)$ 
describe the corresponding wounded-nucleon emission profiles.
Finally,  $f(\eta_\parallel)$ is the emission profile for the binary 
collisions. These functions are chosen appropriately, 
taking into account the following features: the profile $f_+$ ($f_-$) is peaked in the forward (backward) direction, i.e. 
the wounded nucleon emits mostly in its own forward hemisphere, with a quite broad distribution, whereas the binary
profile $f$ is (for the collision of identical nuclei) symmetric. 

We remark that the asymmetric wounded-nucleon emission functions in Eq. (\ref{eq:em}) give a tilt
away from the beam axis for the initial fireball 
(in the $(x,\eta_\parallel)$ plane). The hydrodynamic 
expansion \cite{Bozek:2010bi}
of such a tilted fireball generates the directed flow and 
for the first time the experimental observations  at $\sqrt{s_{NN}}=200$~GeV \cite{Back:2005pc,*Adams:2005ca,*Abelev:2008jga} could be reproduced. 

\begin{figure}[tb]
\includegraphics[width=.44\textwidth]{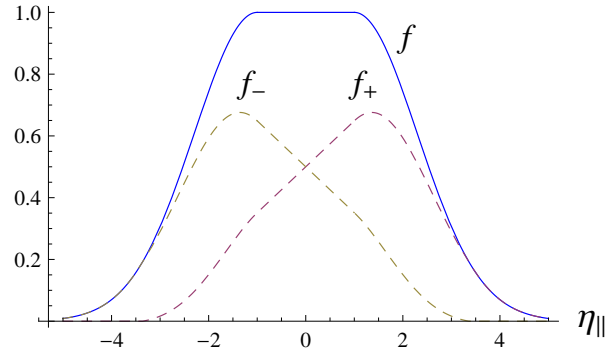}
\caption{(Color online) The emission profiles in space-time rapidity for the wounded nucleons 
(dashed lines) and the binary collisions (solid line). The profile $f+$ ($f_-$) corresponds to the 
forward (backward) moving wounded nucleons. 
\label{fig:profiles}}
\end{figure}

Following Ref.~\cite{Bozek:2010bi}, we choose 
the following parameterizations:
\begin{eqnarray}
f(\eta_\parallel)&=&\exp \left ( - \frac{( |\eta_\parallel| -\eta_0)^2}{2 \sigma_\eta^2} \theta(|\eta_\parallel| - \eta_0)\right ), \nonumber \\
f_+(\eta_\parallel)&=&f_F(\eta_\parallel) f(\eta_\parallel), \nonumber \\
f_-(\eta_\parallel)&=&f_F(-\eta_\parallel) f(\eta_\parallel), 
\end{eqnarray}
with
\begin{eqnarray}
f_F(\eta_\parallel)&=&\left \{ \begin{array}{rr} 0, & \eta_\parallel \le -\eta_m 
           \\  \frac{\eta_\parallel+\eta_m}{2 \eta_m}, & -\eta_m < \eta_\parallel < \eta_m \\1, & \eta_m \le \eta_\parallel  \end{array} \right .
\label{eq:F}
\end{eqnarray}
The values of parameters, describing the RHIC data after 
the hydrodynamic evolution \cite{Bozek:2010bi}, are
\begin{eqnarray}
\eta_0&=&1, \nonumber \\		  
\eta_m&=&3.36, \nonumber \\            
\sigma_\eta&=&1.3.
\label{eq:para}
\end{eqnarray}
The profile functions are shown in Fig.~\ref{fig:profiles}. We note that by 
construction $f_+(\eta_\parallel)+f_-(\eta_\parallel)=f(\eta_\parallel)$.     
Parametrization (\ref{eq:F}) is chosen in such a way, that after the  hydrodynamic evolution and statistical emission \cite{Bozek:2009ty,Bozek:2010bi} 
one correctly reproduces the  spectra \cite{Back:2002wb,*Bearden:2001qq} of particles produced at different rapidities in 
the Au+Au collisions at the highest RHIC energy. 

\subsection{Fluctuations and the torque \label{sec:fluct}}

 \label{sect:glMC}

\begin{figure}[tb]
\includegraphics[width=.25\textwidth]{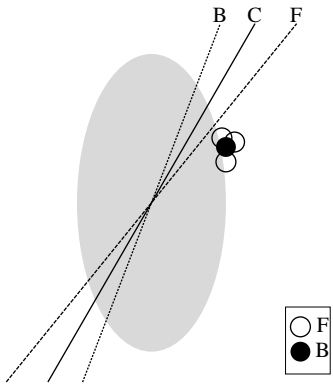}
\caption{A cartoon visualization of the torque effect. A random cluster of wounded nucleons, drawn here at the edge of the ellipse, 
with three nucleons moving in the 
forward (F) direction (open circles) one one moving in the backward (B) direction (filled circle),
causes a random torque of the principal axes. The angle of the torque is higher in the 
forward direction than in the backward direction. In the the central rapidity region (C) the effect is between the F and B 
cases.   
\label{fig:simple}}
\end{figure}

The initial density of the fireball (\ref{eq:em}) can be obtained in a
Glauber Monte Carlo approach. The densities of the forward and backward going 
wounded nucleons  $\rho_\pm(x,y)$ and of the binary collisions $\rho_{\rm bin}(x,y)$ are obtained with 
Monte Carlo simulations by {\tt GLISSANDO} \cite{Broniowski:2007nz}.
These distributions fluctuate on event-by-event basis. 
The phenomenon has a purely statistical 
origin, as the positions on nucleons in nuclei fluctuate.
The event-by-event fluctuations in the wounded-nucleon approach are know
to cause important effects, such as the increase of the elliptic deformation, 
resulting in larger elliptic flow 
\cite{Aguiar:2001ac},  
or the recently discussed triangular flow \cite{Alver:2010gr,*Alver:2010dn,*Petersen:2010cw}. 

Another effect due to fluctuations, focal to this work, is the event-by-event torque of the fireball. Its appearance is simple to 
understand. For the sake of simplicity let us consider the situation depicted in Fig.~\ref{fig:simple}. A cluster of wounded 
nucleons is formed, here drawn at the edge of the fireball. It contains three wounded nucleons moving forward and one moving 
backward. The cluster causes the twist of the principal axis. However, due to the shape of the emission functions of 
Fig.~\ref{fig:profiles}, the shift is different at various values of the space-time rapidity. At forward $\eta_\parallel \sim 3$ there
are practically no backward moving wounded nucleons, hence the three forward-moving nucleons cause the torque. 
In the backward direction only one wounded nucleon from 
the cluster causes the torque, while in the central $\eta_\parallel$ region all four nucleons contribute, but according to Eq.~(\ref{eq:F}) their
relative weight is reduced be a factor of $2$. Thus the relative weight of the effect of the cluster 
in the forward, backward, and central regions is as $3:1:2$.
The result is the schematic arrangement of the principal axes as drawn in Fig.~\ref{fig:simple}.  

In an actual Monte Carlo simulation many clusters occur and the situation is more complicated, but the 
origin is as described above. The effect appears on the event-by-event basis and by symmetry the mean value of the 
torque angle
vanishes upon averaging over events. Thus, the effect  may only be revealed in event-by-event studies of fluctuations, 
see Sect.~\ref{sec:mtf}.

\begin{figure}[tb]
\includegraphics[width=.45\textwidth]{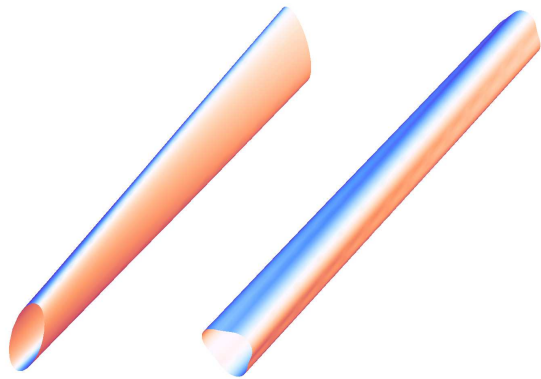}
\caption{(Color online) The schematic figure of the torqued fireball, elongated along the $\eta_\parallel$ axis. 
The direction of the principal axes in the transverse plane 
rotates as $\eta_\parallel$ increases. The left and right pictures correspond to the rank-2 (elliptic) and rank-3 (triangular) cases, 
respectively. The effect occurs event-by-event.
\label{fig:cartoon}}
\end{figure}

A similar phenomenon occurs for the axes of the triangular shape and the axes corresponding to higher Fourier moments in the 
azimuthal angle. The shape of the fireball is depicted schematically in Fig.~\ref{fig:cartoon}, where the torque angle, somewhat exaggerated, 
is shown for the elliptic and triangular deformations. As mentioned above, the torque appears on the event-by-event basis, varying in the 
direction and value.  

\subsection{Characteristics of the torqued the fireball \label{sec:mtf}}

In this subsection we provide quantitative studies of the torque effect at the instant of formation. 
In a given event, the angle of the principal axis for the Fourier moment of rank-$k$ for $n$ sources at some $\eta_\parallel$
is given by the formula 
\begin{eqnarray}
\Psi^{(k)}=\frac{1}{k} {\rm arctan} \left ( \frac{\sum_{i=1}^{n} w_i r_i^2 \sin(k \phi_i)}{\sum_{i=1}^{n} w_i r_i^2 \cos(k \phi_i)} \right ),
\end{eqnarray}
where $(r_i,\phi_i)$ are the polar coordinates of the source position with respect to the center of mass of the slice of the 
fireball at some fixed $\eta_\parallel$, and $w_i$ 
is the weight of the source. 
Explicitly, for the forward-moving wounded nucleons $w_i=(1-\alpha) f_+(\eta_\parallel)$, for the backward ones 
$w_i=(1-\alpha) f_-(\eta_\parallel)$, while for the binary collisions $w_i=\alpha f(\eta_\parallel)$.\footnote{
One could also overlay a statistical distribution of weights, as was done in Ref.~\cite{Broniowski:2007nz}.}
All spatial angles are measured relative to the 
axes perpendicular to the reaction plane. The interpretation of the angle $\Psi^{(k)}$ is that the azimuthal Fourier moment of rank $k$ is 
highest along that direction. The angle is defined modulo $2\pi/k$. For $k=2$ the angle $\Psi^{(2)}$ is the angle of the principal axes of the moment
of inertia. For brevity of notation, we shall skip the superscript $(2)$ from the rank-2 quantities, while retaining superscripts 
indicating  higher ranks.

\begin{figure}[tb]
\includegraphics[width=.47\textwidth]{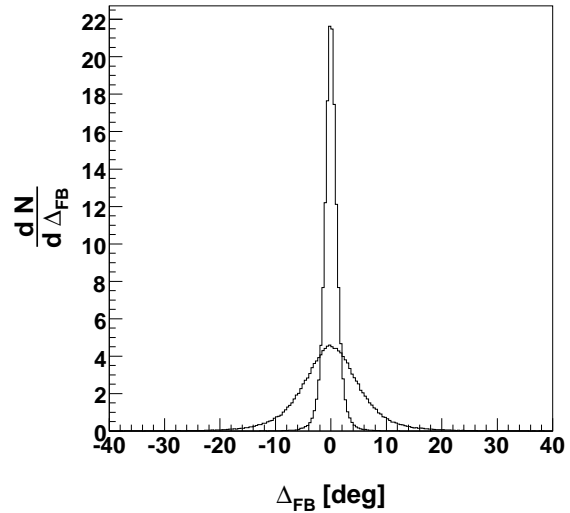} 
\vspace{-7mm}
\caption{
Distribution of the difference of the forward and backward torque angles, $\Psi_F-\Psi_B$, for the elliptic deformation. 
The narrower and wider distributions correspond to the space-time rapidities $\eta_\parallel=0.5$ and $2.5$, 
respectively. Centrality $20-30$\%, mixed model for Au+Au collisions, $\alpha=0.145$.
\label{fig:diffang}}
\end{figure}

The simplest measure of the torque effect is the difference of the $\Psi^{(k)}$ angles at forward and backward 
values of $\eta_\parallel$,
\begin{eqnarray}
\Delta_{FB}^{(k)}(\eta_\parallel)=\Psi^{(k)}(\eta_\parallel)-\Psi^{(k)}(-\eta_\parallel). 
\end{eqnarray}
As argued above, this quantity fluctuates event-by-event. 
The result of the {\tt GLISSANDO} simulations is shown in Fig.~\ref{fig:diffang}. We plot 
the event-by-event distribution of $\Delta_{FB}$ for the 20-30\% centrality class and $\eta_\parallel=0.5$ (the narrower 
distribution) and $\eta_\parallel=2.5$ (the wider distribution). According to the previous qualitative discussion illustrated with 
Fig.~\ref{fig:cartoon}, the widening of the distribution with increasing $\eta_\parallel$ is expected.
The rms width of the distributions of Fig.~\ref{fig:diffang} for $c=20-30$\% are $2^\circ$ for $\eta_\parallel=0.5$ and $7^\circ$ 
for $\eta_\parallel=2.5$.
An analogous study for $c=0-10^\circ$ (not shown) yields $4^\circ$ for $\eta_\parallel=0.5$ and $16^\circ$ for $\eta_\parallel=2.5$. Thus the spread 
enhances with increasing $\eta_\parallel$, moreover, the distributions are wider for the central collisions.  

For the rank-3 analysis (triangular flow) we have qualitatively the same effect.
For $c=20-30$\% the rms width are $4^\circ$ for $\eta_\parallel=0.5$ and $16^\circ$ 
for $\eta_\parallel=2.5$, which are larger than for the rank-2 case at the same centrality.

\begin{figure}[tb]
\includegraphics[width=.5\textwidth]{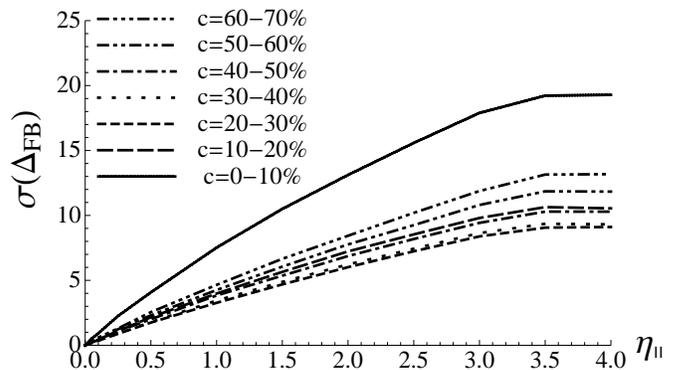} 
\vspace{-1mm}
\caption{The dependence of the rms width of the  $\Delta_{FB}$ distribution on $\eta_\parallel$ at various centralities $c$.
\label{fig:width}}
\end{figure}

\begin{figure}[tb]
\includegraphics[width=.5\textwidth]{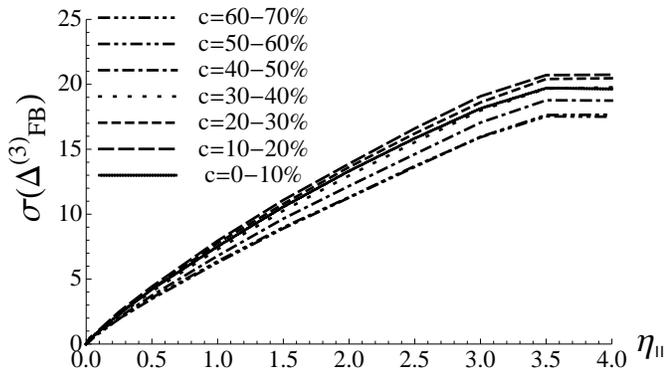} 
\vspace{-1mm}
\caption{Same as Fig.~\ref{fig:width} for the rank-3 case (triangular flow).
\label{fig:width3}}
\end{figure}

In Figs.~\ref{fig:width} and \ref{fig:width3} we show the dependence of the rms
width of the distributions 
of $\Delta_{FB}$ and $\Delta_{FB}^{(3)}$ as functions of $\eta_\parallel$ at various centralities.
For the rank-2 case we note that the widths grow gradually from 0 at $\eta_\parallel=0$ to $10-20^\circ$ at  $\eta_\parallel=4$, with 
the largest angle for the most central events. The rank-3 widths are somewhat larger than in the rank-2 case, except for the most central case.
We note that at fixed $\eta_\parallel$ the dependence on centrality is
non-monotonic, with the lowest value for $c=20-30\%$ for 
the rank-2 case and $40-50\%$ for the rank-3 case. We also note that for the rank-2 case the most central collisions lead to 
significantly larger torque fluctuations than for the less central collisions, cf. Fig.~\ref{fig:width}.

\begin{figure}[tb]
\includegraphics[width=.5\textwidth]{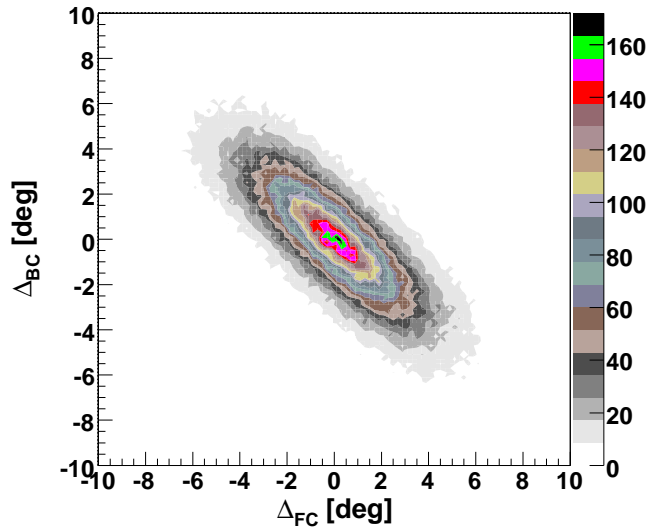} 
\vspace{-2mm}
\caption{(Color online) The 2-dimensional distribution plot of the relative torque angles $\Delta_{\rm FC}$ and $\Delta_{\rm BC}$, 
 for centrality $50-60$\%, space-time 
rapidity $\eta_\parallel=2.5$. The corresponding correlation coefficient is $\rho_{FCB}=-0.61$.
\label{fig:phi2corr2D}}
\end{figure}

One may introduce other measures, involving three angles associated with the forward, backward, and central $\eta_\parallel$ regions.
This might be advantageous, as the central rapidity region is experimentally better covered experimentally.
We define the two relative angles
\begin{eqnarray}
\Delta^{(k)}_{FC}&=\Psi^ {(k)}(\eta_\parallel)-\Psi^ {(k)}(0), \nonumber \\
\Delta^{(k)}_{BC}&=\Psi^ {(k)}(-\eta_\parallel)-\Psi^ {(k)}(0),  
\end{eqnarray}
and their covariance,
\begin{eqnarray}
{\rm cov}^{(k)}_{FBC} =\langle \Delta^{(k)}_{FC}\Delta^{(k)}_{BC} \rangle_{\rm events}. \label{eq:covdef}
\end{eqnarray}
The forward-backward-central correlation coefficient is defined as
\begin{eqnarray}
\rho^{(k)}_{FBC}&\equiv {{\rm cov}^{(k)}_{FBC}}/({\sigma(\Delta^{(k)}_{FC})\sigma(\Delta^{(k)}_{BC})}). \label{eq:corr}
\end{eqnarray}
In Fig.~\ref{fig:phi2corr2D} we present the 2-dimensional distribution plot of $\Delta_{FC}$ and $\Delta_{BC}$ for the rank-2 case for 
$\eta_\parallel=\pm 2.5$ and the 50-60\% centrality class, where the anti-correlation of the angles is clearly 
visible.\footnote{We note that the points in Fig.~\ref{fig:phi2corr2D} occupy all quadrants, i.e. there are cases where both shifts $\Delta_{FC}$ and  $\Delta_{BC}$ have the same sign. 
This is because for each rapidity we evaluate independently the center of mass, which
is the origin for the transverse coordinate system. If we evaluated the angle in the common system 
associated with the central rapidity, only the second and fourth quadrants in Fig.~\ref{fig:phi2corr2D} 
would be filled.} 

\begin{figure}[tb]
\includegraphics[width=.5\textwidth]{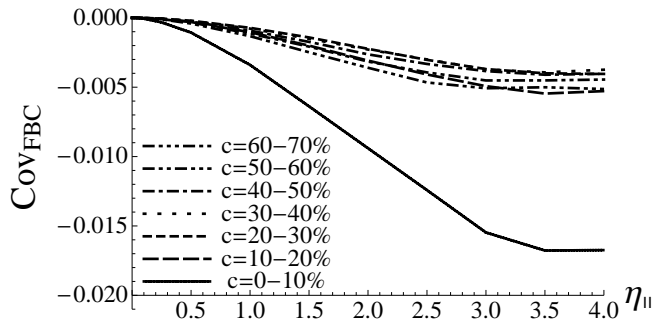} 
\vspace{-2mm}
\caption{Covariance of the $\Delta_{FC}$ and $\Delta_{BC}$ angles plotted as a function of the space-time
rapidity $\eta_\parallel$ for various centralities. 
\label{fig:CovFBC2}}
\end{figure}

\begin{figure}[tb]
\includegraphics[width=.5\textwidth]{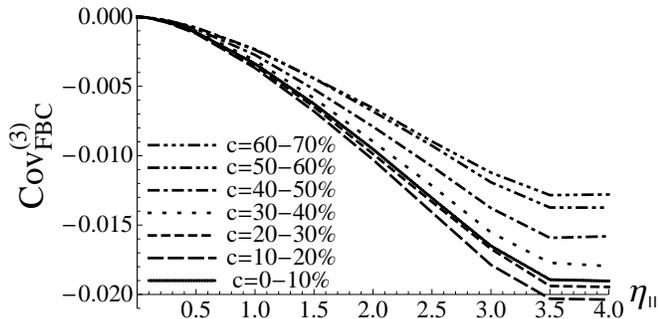} 
\vspace{-2mm}
\caption{Same as in Fig. \ref{fig:CovFBC2} for the rank-3 angles (triangular flow).
\label{fig:CovFBC3}}
\end{figure}

Figures~\ref{fig:CovFBC2} and \ref{fig:CovFBC3} show the rapidity dependence of ${\rm cov}^{(k)}_{FBC}$ for the 
rank-2 and rank-3 cases. We note that these measures drop monotonically with $\eta_\parallel$ from zero  to negative values in the range
$-0.005$ to $-0.02$, reaching a plateau near $\eta_\parallel\approx 3.5$.  
For a fixed $\eta_\parallel$ the dependence on the centrality class is non-monotonic. From the data of 
Figs.~\ref{fig:CovFBC2} and \ref{fig:width} one may obtain the correlation of Eq.~(\ref{eq:corr}). This quantity 
grows from the value  $-1$ at $\eta_\parallel=0$ up to about $-0.5$ at $\eta_\parallel=4$, similarly for the rank-2 and rank-3
cases.

\section{Hydrodynamics \label{sec:hydro}}

The intermediate evolution of the dense system formed in relativistic heavy-ion collisions is believed
to be governed by hydrodynamics 
\cite{Teaney:2000cw,*Kolb:2003dz,*Hama:2005dz,*Huovinen:2006jp,*Hirano:2005xf,*Hirano:2002ds,Broniowski:2008vp,Bozek:2009ty}.  The 
hydrodynamic expansion of the fireball generates a collective velocity
field of the hot fluid. The direction of the acceleration of the fluid element is 
given by pressure gradients. This way the azimuthal eccentricity of 
the fireball is transformed into the elliptic flow 
\cite{Ollitrault:1992bk}, the triangular deformation into the
triangular flow \cite{Alver:2010gr}, and the source tilt 
into the directed flow
of the final hadrons \cite{Bozek:2010bi}. A similar mechanism generates,
on event-by-event basis, a torqued transverse velocity field at different rapidities.

The discussion of the torque of the fireball in the preceding 
sections concerned the earliest stage of the collision, 
described within the wounded-nucleon approach. That stage, essentially, 
prepares the initial condition 
for the subsequent phases of the evolution.
It is necessary to check that the signatures of the torque fluctuations 
survive these later stages, such that measurable effects can be
detected in experiment. Here we investigate the behavior of the rank-2 
torque under the hydrodynamic evolution, 
as the rank-3 case is expected to behave similarly.  A rotation of the
 density in the transverse plane in the 
 torqued fireball scenario, would generate a rotated
 fluid velocity field. This rotation of the transverse 
velocity field at each space-time rapidity would lead to a rotation
 of the reconstructed reaction plane for particles emitted in the corresponding
rapidity range.

We apply the 3+1 hydrodynamic evolution of the perfect fluid with a realistic equation 
of state, implemented for the first time in \cite{Chojnacki:2007jc}. 
This approach is capable of 
uniformly describing the main 
experimental features, such as the transverse momentum spectra, $v_2$,
as well as the Hanbury Brown--Twiss correlation radii \cite{Broniowski:2008vp}. 
Extension to 3+1 dimensions allows to describe also the  spectra at 
non-central rapidities and the directed flow $v_1$ \cite{Bozek:2009ty,Bozek:2010bi}.

To demonstrate the effects of the torqued fireball on final particle spectra, 
we generate the hydrodynamic evolution for the sample value of the impact parameter, $b=6.6$~fm,
which corresponds to the centrality $c=20$-$25$\% \cite{Broniowski:2001ei}. The initial energy density 
$\epsilon$ is taken in the form given in  Eq.~(\ref{eq:em}),
with the wounded nucleon and and binary 
collision densities calculated in the Glauber model. The details and the parameters 
of the distribution can be found in \cite{Bozek:2009ty,Bozek:2010bi}.
The initial proper time for the evolution is $\tau_0=0.25$~fm/c. To study the torque effect, instead of doing  
tedious event-by-event hydrodynamic simulations with a whole distribution of torque
angles as in Fig. \ref{fig:diffang}, we perform one  simulation where 
the densities of the forward and backward going wounded nucleons are rotated 
in opposite directions by a fixed  angle of $5^\circ$ which is a value corresponding to the rms width of the distribution of $\Delta_{FB}$.
Thus the initial energy density 
(\ref{eq:em}), which is the starting point of our hydrodynamics, becomes 
\begin{eqnarray}
&&\epsilon(\eta_\parallel,x,y)=
(1-\alpha)[\rho_+(Rx,Ry) f_+(\eta_\parallel) \nonumber \\
&&+ \rho_-(R^Tx,R^Ty) f_-(\eta_\parallel)] +\alpha 
\rho_{\rm bin}(x,y) f(\eta_\parallel).
\label{eq:emrot}
\end{eqnarray}
The operator $R$ rotates the 
density of forward going wounded nucleons 
by the fixed angle, while  operator $R^T$ rotates the backward going wounded 
nucleons in the opposite direction. 
The density of binary collisions is not rotated, as its emission component 
is symmetric in $\eta_\parallel$.

\begin{figure}[tb]
\includegraphics[width=.45\textwidth]{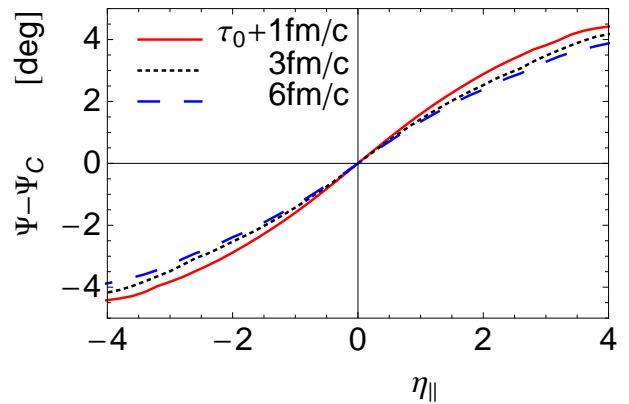} 
\caption{(Color online) The dependence of the
 torque angle of  the fluid velocity field
on space-time rapidity after the 3+1-dimensional hydrodynamics 
of Ref.~\cite{Bozek:2010bi} (solid,  dotted and dashed lines). 
Subsequent curves are for different evolution times.
 \label{fig:hyd}}
\end{figure}

During the  hydrodynamic evolution the 
direction of the transverse flow is determined by the orientation of 
the fireball density at a given space-time rapidity.
 The  torque angle of the fluid, 
 $\Psi(\eta_\parallel)$, is determined by the requirement that in the frame 
defined by the principal axes of the transverse flow we have
$\langle T_{xy} \rangle (\eta_\parallel) = 0$, where 
$\langle T_{\mu \nu} \rangle (\eta_\parallel)=\int dx dy \, 
T_{\mu \nu}(\eta_\parallel,x,y)$ are the components of the energy-momentum tensor 
averaged over the transverse
plane. In Fig. \ref{fig:hyd} we show the evolution of the 
torque angle of the fluid velocity field in the hydrodynamic calculation.
We present the angle as a function of the space-time rapidity after
a hydrodynamic evolution lasting  $1$, $3$ or $6$~fm/c.
Due to the longitudinal push, the effect decreases somewhat as
the evolution time increases, but this quenching is very small. 
Therefore the torque effect survives the hydrodynamic phase with an
almost unchanged magnitude. % , that is  very close to the initial torque angle
% of the fireball (points in Fig. \ref{fig:hyd}). 
A sizable 
twist of the collective velocity field at the freeze-out should 
give a twist of the distribution of the emitted particles.

We note that the space-time rapidity $\eta_\parallel$ is 
not equal to the fluid rapidity, as in a boost-invariant model. Moreover,
the transverse momenta and rapidities of final hadrons include 
a thermal component besides the collective velocity. These effects are
addressed in the next section, through the use of a realistic model of statistical 
hadronization.

\section{Hadronization and experimental measures \label{sec:exp}}

The final question, of key practical importance for the whole idea, 
is how to observe the torque of the fireball from the data consisting 
of momenta of detected particles. In our model approach we adopt the statistical hadronization picture 
\cite{Broniowski:2002nf}, where hadrons (stable and resonances \cite{Torrieri:2004zz}, which subsequently decay) are 
produced at freeze-out according to the Frye-Cooper formalism \cite{Cooper:1974mv}, with a freeze-out temperature of $150$~MeV
\cite{Bozek:2009ty,Bozek:2010bi}.
The difficulty lies in the fact 
that the finiteness of the number of particles produced in each event causes sizable fluctuations of the 
principal axes (or the event plane) as determined from the transverse momenta.  We denote the relative 
angle between the event-plane axis and the fireball spatial principal axis 
$\Psi$ as $\theta$.
Despite this difficulty, 
as we shall see, one can propose measures of even-by-event torque fluctuations that should be possible to be observed in 
the RHIC data. 

Throughout this section $\eta$ denotes the momentum pseudorapidity of produced particles, in distinction 
of the space-time rapidity $\eta_\parallel$ of the preceding parts. 

\subsection{Fluctuations of principal axes from statistical hadronization \label{sec:stat}}

\begin{figure}[tb]
\includegraphics[width=.45\textwidth]{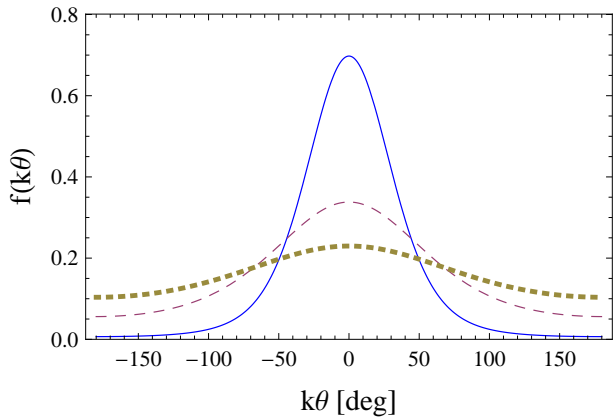} 
\caption{(Color online) The event-by-event distribution of $k\theta$ for $v_k=5\%$ for several values of the 
event multiplicity $n$: 600 (solid), 100 (dashed), and 20 (dotted). \label{fig:angdis}}
\end{figure}

To appreciate the phenomenon of the fluctuation of $\theta$, let us recall formulas from \cite{Broniowski:2007ft} concerning the 
fluctuation of the eccentricity and the principal axes due to the finite number of (independent) particles. 
The angle $\theta$ (dependent on the rank $k$) is defined in each event as 
\begin{eqnarray}
\tan(k \theta)=\frac{\sum_{i=1}^n \sin(k \phi_i)}{\sum_{i=1}^n \cos(k \phi_i)},  
\end{eqnarray}
where all angles are measured with respect to the angle of the principal axis of the fireball, $\Psi^{(k)}$, at a selected $\eta$
window. From Appendix~C of Ref.~\cite{Broniowski:2007ft}, which derives 
 the results based on the central limit theorem,  
we find that (at a fixed multiplicity $n$) 
the event-by-event distribution of $\theta$ is given by
\begin{eqnarray}
&&f(k \theta) = \frac{e^{-{n v_k^2}/{w^2}}}{2 \pi  s^3} \left \{\sqrt{\pi n } v_k \cos (k \theta ) e^{{n v_k^2 \cos ^2(k \theta )}/{(s^2 w^2)}}
 \right . \nonumber \\
&&  \times \left .   \left[\text{sgn}[\cos (k \theta )] \text{erf}\left(\frac{\sqrt{n} v_k |\cos (k \theta )|}{s w}\right)+1\right]+s w\right \}, 
\nonumber \\
\label{eq:clt}
\end{eqnarray}
where the short-hand notation is
\begin{eqnarray}
&& w=\sqrt{1-2v_k^2},  \nonumber   \\
&& s=\sqrt{1-2v_k^2 \sin^2(k \theta)}.
\end{eqnarray}
This distribution is a function of the product $k \theta$ and $v_k \sqrt{n}$,
 therefore we expect universality with 
respect to the rank $k$, as well 
as scaling with $v_k \sqrt{n}$. Both larger multiplicity and larger $v_k$ reduce the fluctuations of the angle $\theta$.

In Fig.~\ref{fig:angdis} we plot the distribution $f(k\theta)$ for the typical value of the flow coefficient for the elliptic flow ($k=2$)
with $v_k=5\%$ for three values of the multiplicity $n$. 
We note a large width of the distributions, of the order of $30^{\circ}$ for $k=2$, or even more as $n$  or $v_2$ decrease. This spread will 
have the tendency
of washing out the smaller torque angle, unless a suitable method, sensitive to differences of angles in the same 
pseudorapidity bin, is used.

\subsection{Cumulants \label{sec:cumul}}

To achieve the goal of observing the torque, similarly to analyses of flow \cite{Ollitrault:1997vz,Borghini:2000sa}, we consider cumulants. 
In the simplest case of the two-particle cumulant we may take
\begin{eqnarray}
\left \langle e^{i n(\phi_F-\phi_B)} \right \rangle = 
\frac{1}{N_{\rm events}} \sum_{\rm events} \frac{1}{n_F n_B} \sum_{i=1}^{n_F} \sum_{j=1}^{n_B}  e^{i k(\phi_i-\phi_j)},  \nonumber \\
\end{eqnarray}
with $k$ denoting the Fourier rank and 
$\phi_i$ ($\phi_j$) being the azimuthal angles of particles emitted in the 
forward (backward) $\eta$ windows in a selected centrality class. The quantities $n_F$ and $n_B$ are the
corresponding multiplicities.  
The measure
 is averaged over events, with the number of events in the sample equal to $N_{\rm events}$.

When no correlations between particles are present, the distribution function of $n$ particles
is the product of one-body distributions. The one-body distribution can then be written in the form 
\begin{eqnarray}
f(\phi) = v_0 + 2 \sum_{k=1} v_k \cos[k (\phi-\Psi^{(k)})].
\end{eqnarray}
Then 
\begin{eqnarray}
\left \langle e^{i k(\phi_F-\phi_B)} \right \rangle = \left \langle v_{k,F} v_{k,B}  \cos (k \Delta_{FB}) \right \rangle_{\rm events},  
\label{eq:cum2}
\end{eqnarray}
where $\left \langle . \right \rangle_{\rm events}$ indicates the averaging over events.
Non-flow contributions modify the right-hand side at the level $1/n$, where $n$ is the effective multiplicity of 
particles in the event.
These effects, hard to estimate, include resonance decays, conservation laws, Bose-Einstein  
correlations of identical particles, short-range correlations, etc. 
The influence of resonance decays will be analyzed via simulations below. 

Since we are interested in measuring the average $\cos [n(\Psi_F-\Psi_B)]$, we need to divide  
Eq.~(\ref{eq:cum2}) by $v_{k,F} v_{k,B}$. We can do it, for instance, by evaluating the ratio of cumulants, defined as 
\begin{eqnarray}
\cos(k\Delta_{FB})\left \{ 2 \right \} &\equiv& \frac{\left \langle e^{i k(\phi_F-\phi_B)} \right \rangle }
{\sqrt{ \left \langle e^{i k(\phi_{F,1}-\phi_{F,2})} \right \rangle \left \langle e^{i k(\phi_{B,1}-\phi_{B,2})} \right \rangle} } = \nonumber\\
&&\hspace{-7mm}\left \langle \cos (k \Delta_{FB}) \right \rangle_{\rm events}  +{\rm nonflow}. \label{eq:c2}
\end{eqnarray}
One may also use higher-order cumulants to generate measures of the torque. For example, 
the ratio of four-particle cumulants, yields
\begin{eqnarray}
\cos(2k \Delta_{FB})\left \{ 4 \right \} &\equiv& 
\frac{\langle  e^{i k [(\phi_{F,1}+\phi_{F,2})-(\phi_{B,1}+\phi_{B,2})]} \rangle}
  {\langle  e^{i k [(\phi_{F,1}-\phi_{F,2})-(\phi_{B,1}-\phi_{B,2})]} \rangle}= \nonumber \\
&&\hspace{-7mm}\left \langle \cos (2 k \Delta_{FB}) \right \rangle_{\rm events}  +{\rm nonflow}
\label{eq:c4}
\end{eqnarray}

The practical issue in this kind of studies is the influence of the non-flow contributions on the result. 
%For the case of resonance decays, in the next section
%we investigate this issue realistically through the use of {\tt THERMINATOR}. 

\subsection{{\tt THERMINATOR} simulations  \label{sec:therm}}

\begin{figure}
\includegraphics[width=.47\textwidth]{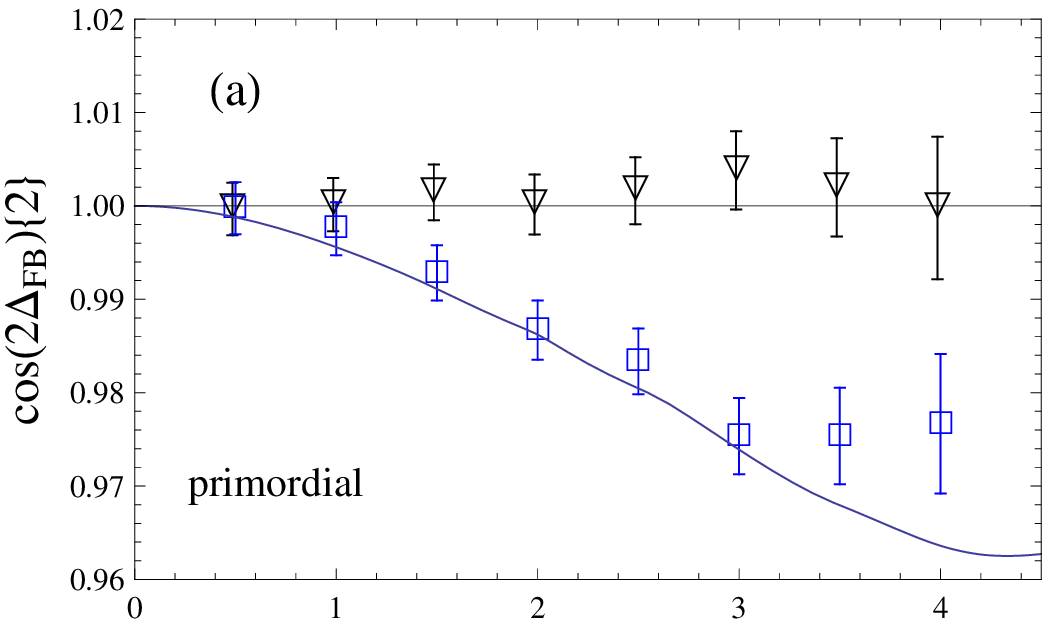} \\ 
\vspace{-8.5mm}
\includegraphics[width=.47\textwidth]{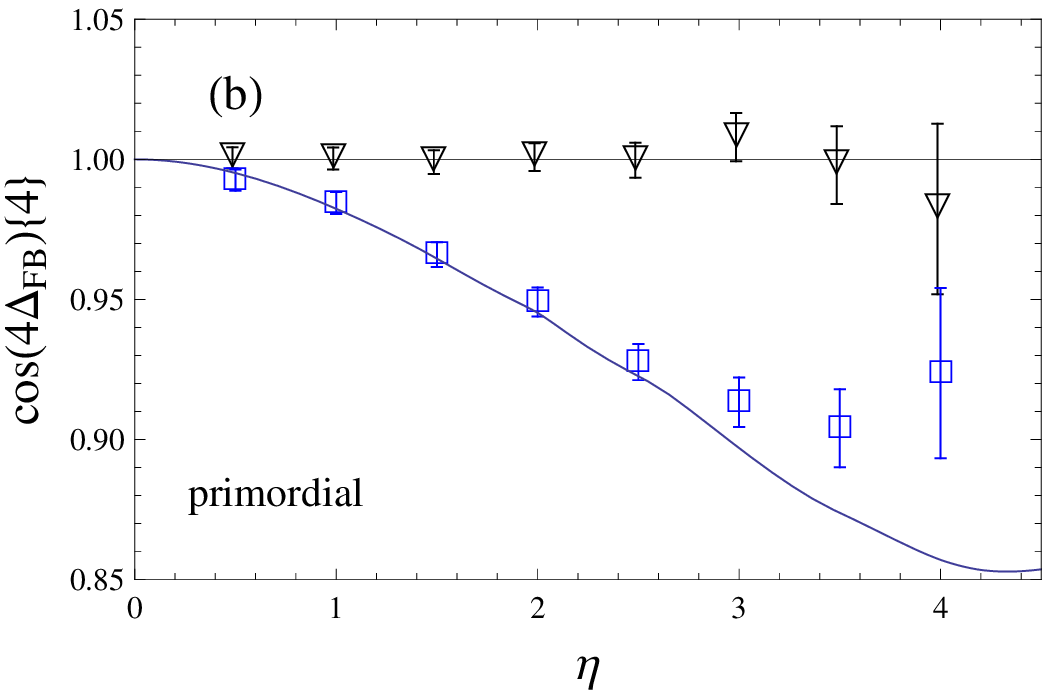} 
\caption{(Color online) Cumulant measures of the torque obtained for the  Au+Au collisions at $c=20-25\%$ 
 with the primordial particles only (i.e. with no resonance decays), plotted as functions of  pseudorapidity. 
Triangles correspond to no torque, squares to the torque of Fig.~\ref{fig:hyd}. 
The solid line represents evaluation directly from   
the fireball torque angle shown in Fig.~\ref{fig:hyd}.  The agreement of the line and the squares 
shows that the statistics is sufficient to detect the torque effect. The $\eta$ windows have the width of one unit. The error bars 
correspond to the statistical errors of the {\tt THERMINATOR} simulation. \label{fig:prim}}
\end{figure}

The precise estimate of the non-flow effects in the formulas of the previous section is not easy. 
To obtain a realistic estimate of the influence of 
resonance decays, we have run {\tt THERMINATOR} \cite{Kisiel:2005hn}, generating 100000 events (in one centrality class) 
from a fireball with a torqued hypersurface, resulting 
from running the 3+1 perfect hydrodynamics \cite{Bozek:2010bi} on the torqued initial condition, as described in Sect.~\ref{sec:hydro}. 
In Figs.~\ref{fig:prim} we present the results 
for centrality $c=20-25\%$, where, at freeze-out, the torque angles in the forward and backward directions differ by  $\simeq 8^\circ$,  
as in Fig.~\ref{fig:hyd}. 

First, we check if {\tt THERMINATOR} is capable of reproducing the (fixed) input
torque angle. For this purpose we take into account primordial particles (those created at the freeze-out hypersurface) only, thus disregarding
resonance decays. The result, shown in Fig.~ \ref{fig:prim} with squares, shows a nice agreement between the cumulant 
measures (\ref{eq:c2},\ref{eq:c4})
and the functions $\cos(2 \Delta_{FB})$ and $\cos(4 \Delta_{FB})$, shown with lines, evaluated directly from   
the fireball torque angle shown in Fig.~\ref{fig:hyd} (we take here the case of the evolution time equal of 6~fm/c).
For comparison, the triangles indicate the result of the calculation without the torque, $\Delta_{FB}=0$.
The agreement shows that without the non-flow contribution from resonance decays  
the accumulated statistics of 100000 {\tt THERMINATOR} events per centrality class is sufficient to see the torque effect.

The error bars in figures of this section correspond to the statistical errors of our Monte Carlo simulation.
These errors increase fast with $\eta$, as the number of produced particles decreases rapidly above $|\eta| \sim 3$.
In experimental data samples, which at RHIC have a very large statistics, these errors should be significantly smaller.

\begin{figure}
\includegraphics[width=.47\textwidth]{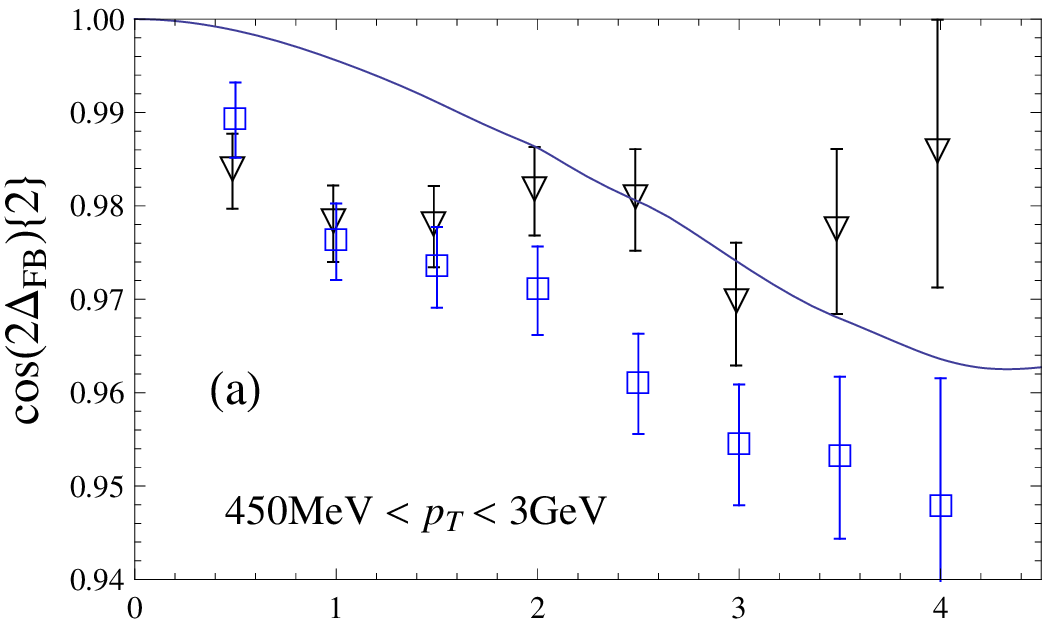} \\ 
\vspace{-5mm}
\includegraphics[width=.47\textwidth]{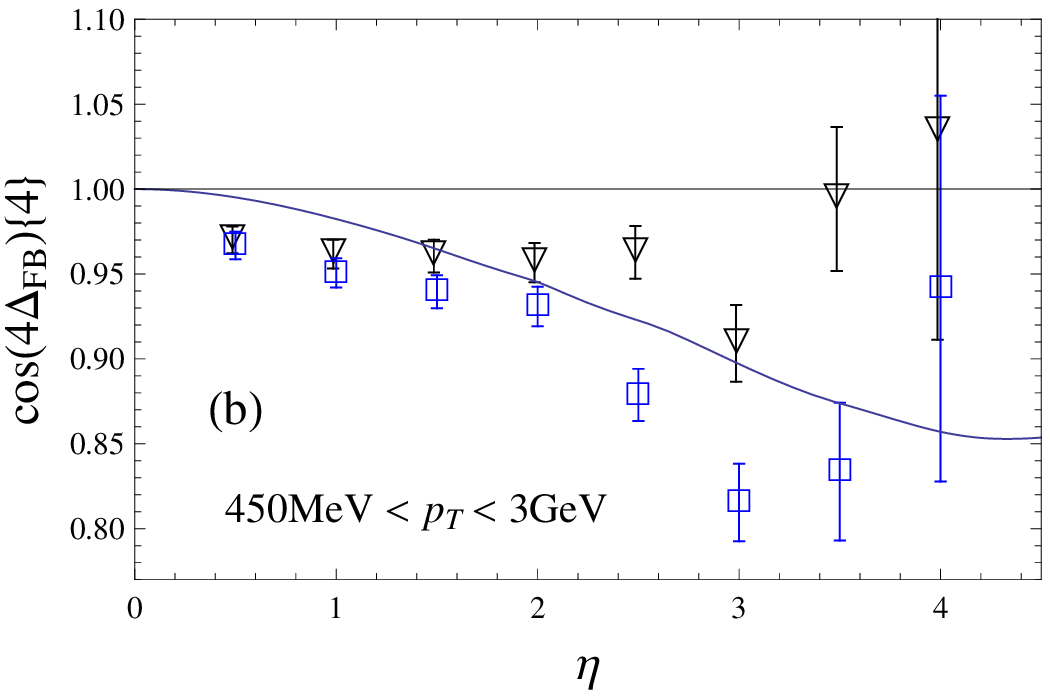} 
\caption{(Color online) Same as Fig.~\ref{fig:prim} obtained from all charged pions, kaons, protons, and antiprotons,
with $450~{\rm MeV} < p_T < 3~{\rm GeV}$. The departure of the triangles, corresponding to no torque, from unity 
displays the non-flow contribution due to resonance decays. The torqued case (squares) is shifted from the 
case without the torque.
\label{fig:above045}}
\end{figure}

Next, we show the realistic case, accounting for all charged pions, kaons, protons, and antiprotons, including those 
coming from resonance decays, in the determination of the 
principal axes. We set the transverse momentum in the window $0.45~{\rm GeV} < p_T < 3~{\rm GeV}$. This limits somewhat the contribution 
of resonance decays, which populate predominantly the softer part of the momentum spectrum. Also, higher $p_T$ leads to a larger elliptic flow
coefficient, which reduces the statistical noise (cf. Eq.(\ref{eq:clt}) and its discussion), 
thus is advantageous for our calculation with a limited-size sample.   
The result of our simulation is shown in Fig.~\ref{fig:above045}, with the squares corresponding to the calculation with the torque 
as in Fig.~\ref{fig:hyd}, and the triangles giving the base-line results with no torque. We note that both 
$\cos(2 \Delta_{FB})\{2\}$ and $\cos(4 \Delta_{FB})\{4\}$ pick up the non-flow effects, as the triangles are shifted down from 
unity.

We see in  Fig. \ref{fig:above045}   that the non-flow effect from resonance decays depends weakly on $\eta$. This is to be expected, as 
the effect comes from correlations of the two particles in the forward pseudorapidity window and the two particles in the 
backward pseudorapidity window. Once these windows are sufficiently separated, the effect does not depend on 
the separation $\eta$. We also remark that the simulation is 
much more noisy (not shown in figures) for the central events, which is due to the small value of $v_2$. 
Thus the torque effect has the best chance of being 
observed for the mid-central or mid-peripheral centrality classes. 
Thus, if a decrease of the proposed cumulant measures with $\eta$ is observed in real data, it would hint 
to the torque effect.

In a complete simulation, an average should be taken of the observables 
over the torque angle distribution. With our calculation, taking the torque angle corresponding to the rms width 
of the angle distribution, we get a correct estimate for the averages of 
quantities  of the order of the square of $\Delta_{FB}$ 
(Eqs. \ref{eq:c2}, \ref{eq:c4}), which is sufficient.

\subsection{Forward-backward-central correlations \label{sec:fbc}}

On may also use correlation measures based on three rapidity windows, forward, backward, and central. One possibility is 
the following combination:
\begin{eqnarray}
&&A_{FBC}\{4\}=\nonumber \\
&&\frac{\langle  e^{i 2 [(\phi_{F}-\phi_{C,1})-(\phi_{B}-\phi_{C,2})]} \rangle - 
\langle  e^{i 2 [(\phi_{F}-\phi_{C,1})+ (\phi_{B}-\phi_{C,2})]} \rangle}
{v_{2,F} v_{2,B} v^2_{2,C}} \nonumber\\
&&= \langle 2 \sin(2 \Delta_{FC}) \sin(2 \Delta_{BC}) \rangle_{\rm events} +{\rm nonflow}.
\label{AFBC}
\end{eqnarray}
For small torque angles this measure reduces to the covariance (\ref{eq:covdef}), namely
\begin{eqnarray}
A_{FBC}\{4\} \sim 8 \,{\rm cov}(\Delta_{FC},\Delta_{BC}) +{\rm nonflow}.
\end{eqnarray}

\begin{figure}[tb]
\includegraphics[width=.47\textwidth]{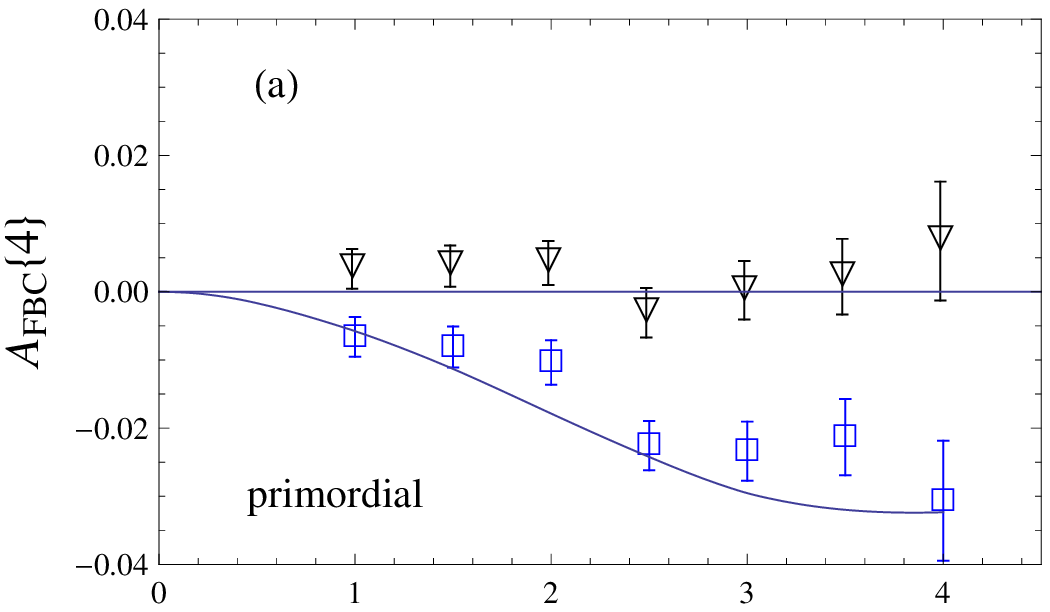}\\
\vspace{-5mm}
\includegraphics[width=.47\textwidth]{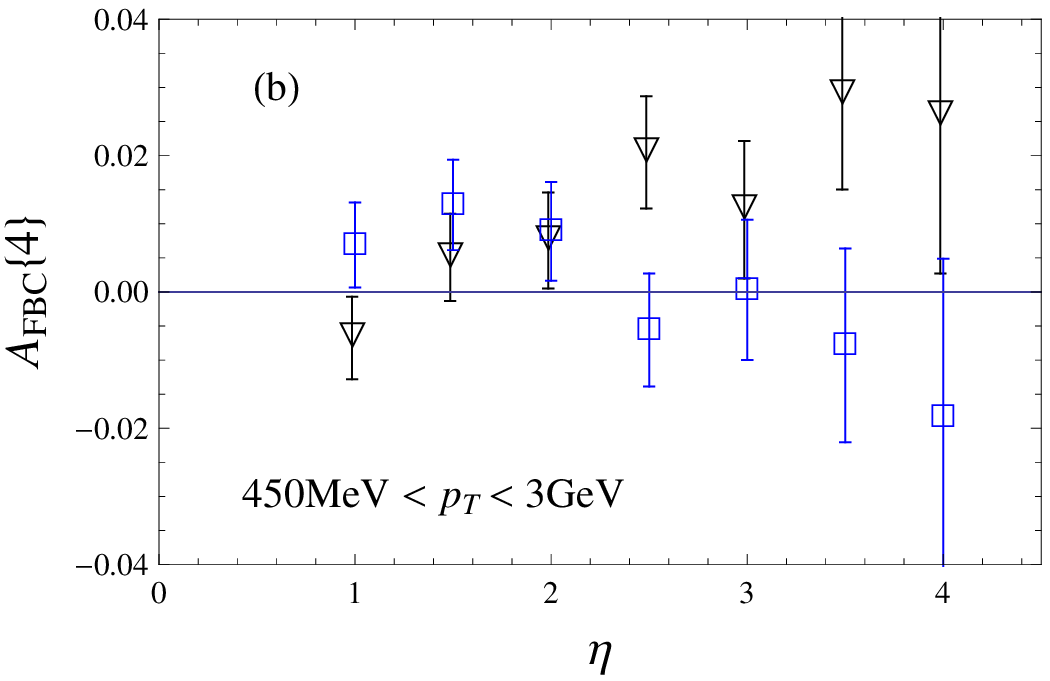}
\vspace{-1mm}
\caption{The measure $A_{FBC}$ obtained from for the primordial particles (a) and all charged 
pions, kaons, protons, and antiprotons (b).
Triangles correspond to no torque, squares to the torque of Fig.~\ref{fig:hyd}. The solid line shows 
$8 {\rm cov}_{FBC}$ of Eq.~(\ref{eq:covdef}). The $\eta$ windows have the width of one unit. The error bars 
correspond to the statistical errors of the {\tt THERMINATOR} simulation.
\label{fig:AFBC}}
\end{figure}

The results of the {\tt THERMINATOR} simulations for $A_{FBC}$ are shown in Fig.~\ref{fig:AFBC}. The conclusions are similar as for the 
previously discussed measures. With primordial particles (no non-flow effects, Fig.~\ref{fig:AFBC}(a)) we reproduce 
the expected behavior.  The squares are very close to the solid line, showing $8 {\rm cov}_{FBC}$
(cf. Fig.~\ref{fig:CovFBC2}). For the 
realistic case of all charged pions, kaons, protons, and antiprotons (Fig.~\ref{fig:AFBC}(b)), 
we observe the non-flow effect from the resonance decays. We note that 
the cases of the torqued and untorqued fireballs are qualitatively different, with the former giving positive, and the latter negative values 
of $A_{FBC}$.   

\section{Conclusions \label{sec:conclude}}

In conclusion, we summarize our results:

\begin{enumerate}

\item The space-time rapidity emission profile, where the wounded nucleons emit predominantly in the 
direction of their motion, combined with the statistical fluctuations of the source densities in the transverse plane, 
lead to event-by event torqued fireballs.

\item The standard deviation of the relative torque angle between the forward ($\eta_\parallel \sim 3$) 
and backward ($\eta_\parallel \sim -3$) regions varies from $20^\circ$ for the most central collisions to $10^\circ$ 
for the mid-central and  mid-peripheral Au+Au collisions at the highest RHIC energies.

\item  This initial torque is transformed, via hydrodynamics, into the torque 
of the transverse collective flow of the fluid, and subsequently into the torque of the 
principal axes of the transverse-momentum distributions of the detected particles. 

\item Statistical measures based on cumulants containing particles in different pseudorapidity bins 
should be useful in detecting the torque effect experimentally. 

\item The non-flow corrections can be sizable, but should not overshadow the effect. 
In particular this is the case of the effects of resonance decays, estimated realistically with {\tt THERMINATOR} Monte Carlo 
simulations. We find that there is a clear difference between the behavior of the proposed cumulant measures for the untorqued and 
torqued cases. Therefore the torque fluctuations should  possibly  be 
observed in the high-statistics RHIC data by the PHOBOS and STAR collaborations.

\item Since the statistical noise increases as the product of the particle 
multiplicity and the $v_k$ flow coefficient, the best chance for looking for the 
torque effect is in the mid-central or mid-peripheral centrality classes, such as $c=20-30\%$, 
and with the exclusion of the soft-momentum hadrons.

\item Based on our calculations, the torque is expected to have a 
similar size for the elliptic flow and triangular flow.

\item Other effects that may influence the torque effect, such as the conservation of
momentum \cite{Borghini:2006yk,*Becattini:2007nx,*Hauer:2007ju,*Hauer:2007im,*Chajecki:2008yi,*Bzdak:2010fd}, 
angular-momentum \cite{Becattini:2007sr}, 
or charge \cite{Jeon:2000wg,*Koch:2008ia}, as well as other sources 
of correlations and fluctuations, should be incorporated in future studies. 

\item Finally, the torque fluctuations
in other two-source or multi-source models 
\cite{Bialas:2007eg,*Andersson:1986gw,*Capella:1992yb,*Lappi:2006fp,*Gelis:2009wh} 
should be investigated.

\end{enumerate}

\begin{acknowledgments}
Supported by Polish Ministry of Science and Higher Education, grants N~N202~263438 and N~N202~249235, and by 
the Portuguese Funda\c{c}\~{a}o para a Ci\^{e}ncia e Tecnologia, FEDER, OE, grant SFRH/BPD/63070/2009, CERN/FP/116334/201.
\end{acknowledgments}

\bibliography{hydr}

\end{document}